\newcommand{\f}[1]{Fig.~\ref{#1}}
\newcommand{\eq}[1]{Eq.~(\ref{#1})}

\def\p{\protect}
\def\d{\partial}
\def\be{\begin{equation}}
\def\ee{\end{equation}}
\def\bea{\begin{eqnarray}}
\def\eea{\end{eqnarray}}
\def\l({\left(}
\def\r){\right)}

\newcommand{\y}{Y\-Ba$_{2}$\-Cu$_{3}$\-O$_{7-\delta}$}
\def\s2{w/\sqrt{2}}

\documentstyle[prb,aps,multicol,psfig]{revtex}

  \renewcommand{\narrowtext}{\begin{multicols}{2} \global\columnwidth20.5pc}
  \renewcommand{\widetext}{\end{multicols} \global\columnwidth42.5pc}

\begin{document}
\title{Symmetry of the remanent state flux distribution  in
superconducting thin strips: Probing 
the critical state}
\author{A. V. Bobyl$^{1,2}$, 
D.~V. Shantsev$^{1,2}$,
Y.~M.~Galperin$^{1,2}$, T.~H.~Johansen$^{1,}$\cite{0}}
\address{
$^1$Department of Physics, University of Oslo, P. O. Box 1048
Blindern, 0316 Oslo, Norway\\
$^2$A. F. Ioffe Physico-Technical Institute, Polytekhnicheskaya 26,
St.Petersburg 194021, Russia
}
\date{\today}
\maketitle

\begin{abstract}
The critical-state in a thin strip of \y\
is studied by magneto-optical imaging.
The distribution of magnetic flux density is 
shown to have a specific symmetry in the remanent state
after a large applied field.
The symmetry was predicted [\prl {\bf 82}, 2947 (1999)]
for any $j_{c}(B)$,
and is therefore suggested as a simple tool to verify
the applicability of the critical-state model. 
At large temperatures we find deviations from 
this symmetry, which demonstrates departure from the 
critical-state behavior.
The observed deviations can be attributed to an explicit
coordinate dependence of $j_c$ since both a surface 
barrier, and flux creep would break the symmetry in a different way.
\end{abstract}

\narrowtext

\section{Introduction}

During the last years much attention has been paid to studies of
the magnetic behavior of thin superconductors placed in a 
perpendicular magnetic field, the so called perpendicular geometry.
On one hand, numerous papers have investigated theoretically the
critical state of thin superconductors of regular shapes like long 
strips and circular disks.\cite{BrIn,zeld,Mik,Zhu,Clem,Mikitik,McD,diskjcb}  
On the other, magnetic characterization by measuring the spatial distribution 
of flux density at the surface of superconductors has become quite
common and powerful. This progress has been facilitated by the
development of spatially-resolved techniques, such as 
magneto-optical imaging, Hall-probe arrays etc., see
Ref.~\onlinecite{Brandt} for a review.

In spite of these massive efforts, the task of making proper
interpretations of  
a measured flux density distribution, $B({\bf r})$, is still one with
major difficulties. 
In particular, we are not aware of any simple decisive method to judge whether an observed
$B({\bf r})$ is consistent with the critical-state model (CSM) or not. 
One could here expect that fitting an observed 
$B({\bf r})$ by profiles predicted from the CSM with some $B$-dependent
critical current density, $j_c(B)$, would be a straightforward procedure.
However, this is not so since in the perpendicular geometry
{\em explicit} expressions for the flux density distribution are available only for the Bean
model, $j_c=\text{const}$, for a thin strip\cite{BrIn,zeld} and a thin disk\cite{Mik,Zhu,Clem}, 
 as well as for a strip with a special kind of $j_c(B)$.\cite{Mikitik} 
For a thin strip or disk with a general $B$-dependent $j_c$, the flux distribution  
can be calculated only numerically by solving a set of integral equations.\cite{McD,diskjcb}   

In a recent work,\cite{PRL} we considered the critical-state magnetic behavior 
of a thin superconducting strip with a general $B$-dependence of $j_c$.
It was shown that the central peak in large-field magnetization hysteresis loops
of such samples always occurs at the remanent state, $B_a=0$.
An intermediate result of that derivation is the prediction
of a special symmetry of the remanent-state flux density distribution.
Since the symmetry is independent of the particular $j_c(B)$, it may serve as 
a conclusive and easily implementable test for the applicability of the critical-state 
model in a given experiment. In the present paper we
demonstrate using magneto-optical imaging how this symmetry in the flux density profile
can be revealed, and used to verify the applicability of the CSM.

\section{Symmetry of flux distribution}

Consider a long thin superconducting strip with edges located at $x=\pm w$,
the $y$-axis pointing along the strip, and the $z$-axis normal to the strip
plane, see \f{f_strip}. The magnetic field,  $B_{a}$, is applied along
the $z$-axis, so screening currents are flowing in the
$y$-direction. Throughout the paper
$B$ denotes the $z$-component of magnetic induction in the strip plane. The sheet
current is defined as $J(x)=\int {j(x,z)\, dz}$, where $j(x,z)$ is the current
density and the integration is performed over the strip thickness, $d\ll w$.

\begin{figure}[tbp]
\centerline{ \psfig{figure=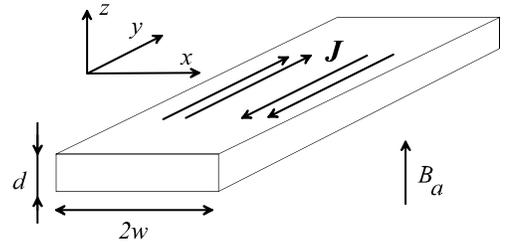,width=8cm}}
\caption{Superconducting strip in a perpendicular applied magnetic field. }
\label{f_strip}
\end{figure}

{}From the Biot-Savart law for the strip geometry, the flux density is
given by~\cite{BrIn,zeld}
\begin{equation}
B(x)-B_{a}=-\frac{\mu _{0}}{2\pi }\int_{-w}^{w}\frac{J(u)\,du}{x-u}\, .
\label{B}
\end{equation}
Assume that the strip is in the remanent state after a very large field
was applied. Then, everywhere in the strip the
current density is equal to the critical value,
$J(x)={\rm sgn}(x)\,J_{c}[B(x)]$, and the flux density
distribution satisfies the following integral equation,
\begin{equation}
B(x)=-\frac{\mu _0}{\pi } \int_{0}^{w}
  \frac{J_c[B(u)] }{x^2-u^2}\,u\, du\, .
\label{ie}
\end{equation}
By changing here the integration variable
from $u$ to $\sqrt{w^{2}-u^{2}}$, and also replacing $x$ by $\sqrt{w^2-x^2}$, 
one obtains
\begin{equation}
B(\sqrt{w^2-x^2})=\frac{\mu _0}{\pi }\int_{0}^{w}
\frac{J_c [B(  \sqrt{w^2-u^2}  )] }
{x^2-u^2} \,u \, du \,.
\label{tm4}
\end{equation}
Clearly, this equation for $-B(\sqrt{w^{2}-x^{2}})$
is equivalent to \eq{ie} for $B(x)$ if $J_{c}$ depends only on the absolute value of
the magnetic induction, $J_{c}(|B|)$. Thus, we conclude that
\begin{equation}
B(x)=-B(\sqrt{w^{2}-x^{2}})\, .
\label{tm5}
\end{equation}
This symmetry of the flux density distribution in the fully-penetrated remanent state
is valid for {\em any} $J_{c}(|B|)$ dependence. In particular, it also holds for 
the Bean model, where one finds by simple integration of \eq{ie} that
\begin{equation}
B(x)=\frac{\mu _0 J_c}{2 \pi } \, \ln \frac{w^2 - x^2}{x^2} \ .
\label{bean}
\end{equation}
It follows from the general \eq{tm5} that the flux density is always zero at $x=w/\sqrt{2}$.
At this point $B(x)$ changes sign from positive in the central part of the strip to
negative near the edges.

This special symmetry of the flux density profile across a thin strip has a trivial analog
for the case of a long sample in a parallel field. There, the gradient in $B$ is a 
function of the local value of the field. Hence, the flux distribution is always
symmetric around the point $x_0$ where the flux density is zero, $B(x_0)=0$. Hence, 
one can write $B(x)=-B(2x_0-x)$ as long as $x$ and $2x_0-x$ are within the superconductor.
In the perpendicular geometry the relationship between $B$ and $j$
is non-local, and the $B$-profiles deviate strongly from those in the 
interior of long slabs and cylinders.

\section{Experiment}

A 200~nm thick film of \y\ was grown epitaxially on an MgO substrate
using laser ablation.\cite{Shen} The sample was patterned by chemical etching 
into a strip shape of width $2w=2.5$~mm. For the measurements of flux density profiles
we chose a region free of any defects visible by our magneto-optical (MO) imaging
system.

\begin{figure}
\centerline{ \psfig{figure=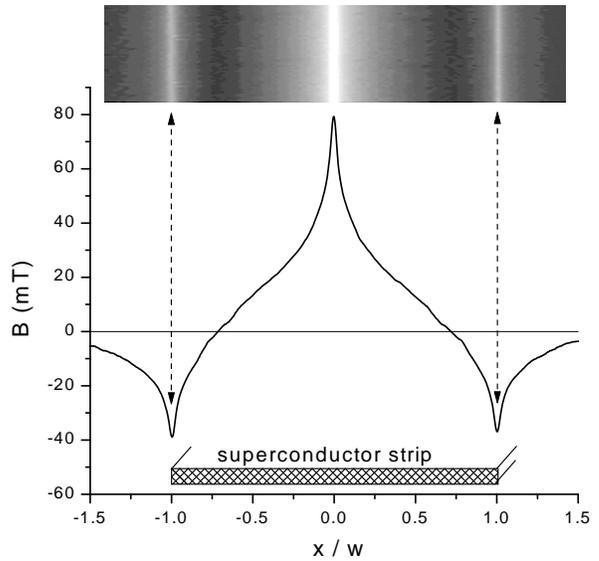,width=8cm}}
\caption{Flux density profile across the \y\ strip in the remanent state at 42~K.
The profile is derived from the magneto-optical image shown above.
The bright areas on the image correspond to large $|B|$.
\label{f_mo}}
\end{figure}
\begin{figure}
\centerline{ \psfig{figure=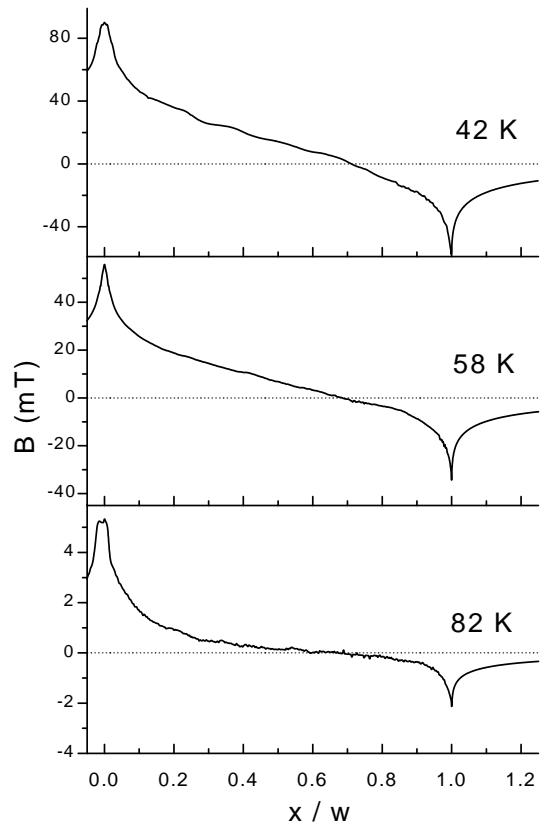,width=8cm}}
\caption{Profiles of the flux density across one half of the \y\ strip at
different temperatures. The strip is in the remanent state after a large field
had been applied.
\label{f_bj}}
\end{figure}
The imaging system is based on the Faraday rotation of
polarized light illuminating an MO-active indicator film that we
mount directly on top of the superconductor's surface. The rotated 
Faraday angle varies locally with the value of the perpendicular magnetic 
field, and 
through a crossed analyzer in an optical
microscope one can directly visualize and quantify the field
distribution across the covered sample area. As Faraday-active indicator we
use a  Bi-doped yttrium iron garnet film with in-plane magnetization.\cite{dor1} 
The indicator film was deposited to a thickness
of 5 $\mu$m by liquid phase epitaxy on a gadolinium gallium garnet
substrate. A thin layer of aluminum is evaporated onto the
film allowing incident light to be reflected, thus providing 
double Faraday rotation of the light beam. The images were recorded
with an 8-bit Kodak DCS 420 CCD camera  and transferred to a
computer for processing. The conversion of gray levels into
magnetic field values is based on a careful calibration, see
Ref. \onlinecite{Joh96}.
The MO imaging at low temperatures was performed by mounting the 
superconductor/MO-indicator on the cold finger of a continuous 
He-flow cryostat with an optical window (Microstat, Oxford).

MO images were taken in the remanent state after applying a large field 
at temperatures of 42, 58 and 82~K.
An MO image at 42~K and the corresponding flux density profile are shown in \f{f_mo}.
The fact that maximum trapped flux density
is observed in the center of the strip, implies that the applied field had been raised to 
a sufficiently large value. Furthermore, one sees from the figure that the return field of 
the trapped flux penetrates regions near the strip edges, $x=\pm w$,
where the field has negative peaks. 
While the intensity in the MO image does not immediately discriminate between the two 
field polarities, it is readily accounted for by locating the boundary where $B=0$, also called 
the annihilation zone. 
One sees also that the flux distribution in the left and right halves of the strip are mirror 
images of each other, and therefore we focus only on one half of the strip, $0\le x \le w$.
The flux density distribution at higher temperatures are shown in \f{f_bj}.
Due to reduced flux pinning with increasing temperatures the magnitude of the trapped
field is reduced substantially. In addition, we also see changes in the shape of the flux 
profile.
The spatial resolution of the method is limited by the thickness of the MO indicator film.
Therefore, a few data points in 5~$\mu$m vicinity of the peaks at the strip center and at the edge
have been removed from the following analysis. 

To test the symmetry property, expressed in \eq{tm5}, of the measured flux profiles
we plot in \f{f_sym} the absolute of the flux density as a function of the new 
coordinate $x'$,
\be
  x' = \left\{
  \begin{array}{ll}
  x, & 0<x \le w/\sqrt{2}\\
  \sqrt{w^2-x^2}, & w/\sqrt{2} < x < w \ .
  \end{array}
  \right.
\label{x'}
\ee
If \eq{tm5} holds, there should be full overlap of the two branches
where the measured $B$  
is positive and negative.
One sees from the figure that the overlap is almost complete for the
data taken at 42 and 58~K except for small deviations at large $|B|$.
At 82~K, however, there is a significant splitting of the
two branches over the whole range. 
One may therefore conclude that there is a systematic deviation
from the CSM behavior at this high temperature. 
\begin{figure}
\centerline{\psfig{figure=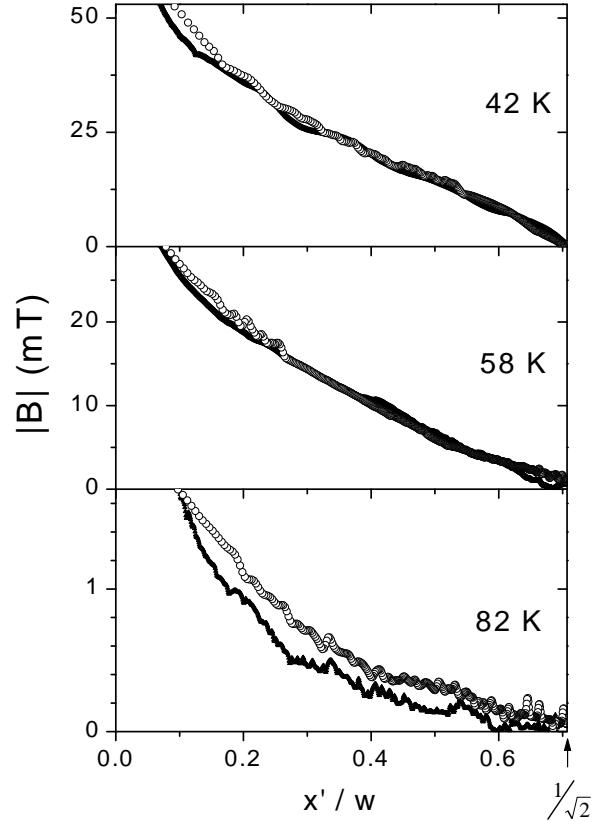,width=8cm}}
\caption{Profiles of the flux density from \p\f{f_bj}
replotted with new coordinate $x'$ defined by \p\eq{x'}. 
Open and solid symbols correspond to $x>w/\p\sqrt{2}$,
and $x<w/\p\sqrt{2}$, respectively.
A large splitting of the two branches at 82~K indicates
a deviation from the CSM.}
\label{f_sym}
\end{figure}

The results from the symmetry analysis of
$B$-profiles are now compared with direct evaluation of  
$j$ in regions with equal $|B|$. 
For this purpose the current density distributions were calculated
from the measured $B$-profiles by the inversion scheme
proposed in Ref.~\onlinecite{Joh96} 
and developed elsewhere.\cite{future} The latter procedure is, in
general, much more complicated than the analysis of $B$-profiles and
it requires knowledge of $B$-distributions across rather large regions
also outside the strip. Furthermore, the inversion procedure involves
filtering of short-wavelength noise in experimental data. 

The current profile at 42~K is shown as a solid line in the inset of \f{f_jc}. 
One can see an enhancement in $j$ in the region
near $B=0$. In the main figure the data
are replotted as $j$ versus $|B|$.
Again we see that the two branches corresponding to positive and 
negative $B$ collapse, which proves existence 
of the critical state in the strip. 
The unified curve characterizes the $j_c(B)$ dependence.
Also in this plot, like in \f{f_sym}, there are small 
deviations from  data collapse at the largest fields. 
We thus see that the analysis based on $j$-inversion gives
similar results as the symmetry analysis of $B$ profiles.

\begin{figure}[tbp]
\centerline{ \psfig{figure=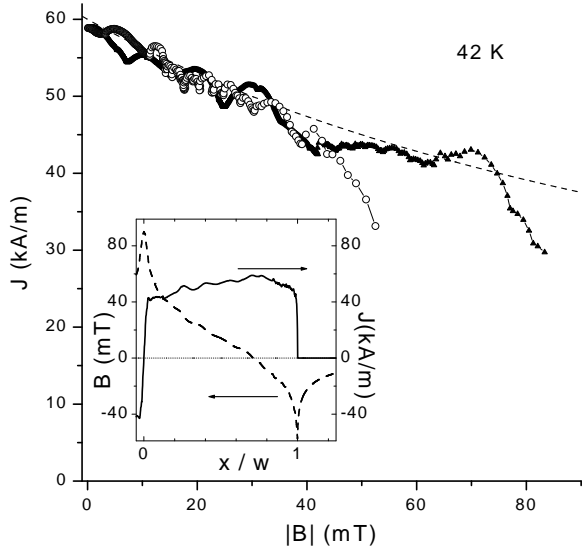,width=8cm}}
\caption{Sheet current versus the absolute value of the local flux density at 42~K.
The data are obtained by replotting the $B(x)$ and $j(x)$ profiles shown in the inset.
A collapse of the two branches, which correspond to positive and negative $B$ (solid and open symbols),
demonstrates that the critical-state is established in the strip.
The dashed line shows a fitted Kim model $j_c(B) \propto (1+B/B_0)^{-1}$ 
with $B_0=150$~mT.
\label{f_jc}}
\end{figure}

\section{Discussion}

The symmetry condition, \eq{tm5}, was derived within the framework of the CSM, 
and is not expected to hold when one or more of the basic assumptions of this
very successful model is violated.
The most probable reasons for lack of this symmetry are therefore the following; 

(i) Presence of a {\em surface barrier} for vortex entry  and exit. 
There are vast experimental observations of such a barrier in high-$T_c$ superconductor
crystals. The barrier leads to an excessive current density in the vicinity of 
edges, which then destroys the symmetry. 
In particular, it shifts the point $x_0$ where $B=0$ 
towards the strip edge.
For the geometrical barrier which arises from a rectangular shape of the cross sectional area,
the excessive current density is of the order of $H_{c1}/d$.\cite{ZeldovGB}
For the Bean-Livingston barrier it should be larger and scale with temperature as the
thermodynamic critical field, $H_{c}$.\cite{BL} 
In both cases, the temperature dependence can differ 
from that of the bulk pinning $j_c$ suggesting that deviation from the symmetry 
can be temperature dependent.
Moreover, it is known from experiment\cite{FuchsNature} that the surface barrier 
dominates at the higher temperatures, while bulk pinning is more
important at low temperatures. 

(ii) {\em Thermally-activated creep} of vortices leading to a slow time
relaxation of the flux distribution. 
The flux creep problem for a thin strip with 
a $B$-independent  $E(j)$ law
has been considered in Refs.~\onlinecite{Creep-strip,brandt-uni}.
Under constant applied magnetic field the space and time dependence of 
the electric field 
is shown to decouple as $E(x,t)=f(x) g(t)$, where 
$f(x)$ can be found numerically. It is also argued
that during relaxation of $E(x,t)$, starting from some initial $E(x,0)$,
the electric field will approach the profile given by $f(x)$.
From $\dot{B}=-\d E/\d x$ it follows that 
if an initial remanent flux profile $B(x)$ crossed zero at $x_0=w/\sqrt{2}$,
then during relaxation $x_0$ will shift towards the point
where $f(x)$ has the maximum.
For the voltage-current law $E=E_c (j/j_c)^n$, the maximum in $f$ 
is always located at $x_0>w/\sqrt{2}$, namely 
at $x_0=0.735w$ for $n=1$ and approaches $w/\sqrt{2}$ as $n \rightarrow \infty$.\cite{brandt-uni}
This means that at smaller $n$, i. e., at larger temperatures the deviations
from the symmetry due to relaxation are stronger.\cite{note}

Thus, both a surface barrier as well as flux creep predict a stronger 
deviation from the symmetry in $B(x)$ at higher temperatures. 
However, neither of them can explain the deviation 
found in our experiment.
Indeed, while the CSM predicts that in the remanent state $x_0=w/\sqrt{2}$,
both surface barrier and flux creep lead to larger $x_0$. 
Such a shift of $x_0$ would result in the negative-$B$ branch being below the
positive-$B$ branch in the  $|B|(x')$ plot.
However, that is just the opposite to what is shown in \f{f_sym}.

(iii) The last possible reason for the deviation from the symmetry
is  {\em inhomogeneity of the strip},
which leads to an explicit coordinate dependence of the 
critical current density, $j_c(x,B(x))$. 
It may be caused, e. g., by a nonuniform chemical composition.\cite{JAP}
The kind of deviation shown in \f{f_sym} can be explained 
by a suppressed $j_c$ near the strip edge.
The fact that strong deviations are found only at the highest temperature
can be related to the existence of two mechanisms controlling $j_c$ with
different temperature dependences. If so, only the mechanism dominant at high $T$
has to produce an inhomogeneous $j_c$. An example of two such mechanisms
can be bulk and inter-grain pinning, which are known to have 
different $T$-dependences.\cite{welp,polyanskii,coup} 
In thin \y\ films the second mechanism can be realized on any planar defect
such as a boundary between microblocks with slightly different crystal axis orientation,\cite{jap97}
a twin boundary, or a microcrack.\cite{wordenweber} 

\section{Conclusions}

The flux density distribution in a
superconducting thin strip with a general
$j_{c}(B)$ is shown to have a special kind of symmetry
in the remanent state after large applied field.
Probing the symmetry of measured flux distributions is suggested as a
simple method to test 
applicability of the critical-state model without {\em a priori}~ knowledge of 
$j_{c}(B)$. The procedure is simpler than calculation of the
current distributions because it requires knowledge only of the field
inside the strip and it is also weakly sensitive to ``noise'' in the
experimental data. 
The method has been applied to a thin \y\ strip which exhibited a fairly good CSM behavior
well below $T_c$, but large deviations from the symmetry were observed at 82~K.
Our analysis shows that the deviations can be attributed to an explicit
coordinate dependence of $j_c$ since both a surface 
barrier, and strong flux creep would break the symmetry in a different way.


The financial support from the Research Council of Norway (NFR), and from
NATO via NFR is gratefully acknowledged. We are grateful to Bj{\o}rn Berling for a 
many-sided help.


\widetext

\begin{references}
\bibitem{note}Remarkably, for the special dependence
$E =E_c \exp\l(j/j_c\r)$ with constant $j_c$,
the flux creep does not affect the symmetry at all. In this case\cite{Creep-strip} 
$
   f = (1+\eta) \ln(1+\eta) - (1-\eta) \ln(1-\eta) - 2 \eta \ln|\eta| 
$
with $\eta=x/w$, which means that
the derivative $df/dx$ has exactly the same symmetry as the field profile, \eq{tm5}.
This case, however, does not describe our experiment 
where $J_c$ is obviously $B$-dependent.
\end{references}
\end{document}